\documentclass[sigconf,screen]{acmart}
\acmConference[ESEM 2024]{The 18th ACM/IEEE International Symposium on Empirical Software Engineering and Measurement}{20–25 October, 2024}{Barcelona, Spain}
\AtBeginDocument{%
  \providecommand\BibTeX{{%
    \normalfont B\kern-0.5em{\scshape i\kern-0.25em b}\kern-0.8em\TeX}}}

\setcopyright{acmcopyright}
\copyrightyear{2024}
\acmYear{2024}
\acmDOI{XXXXXXX.XXXXXXX}

\usepackage{amsmath,amsfonts}
\usepackage{algpseudocode}
\usepackage{graphicx}
\usepackage[many]{tcolorbox}
\usepackage{textcomp}
\usepackage{xcolor}
\usepackage{color, colortbl}
\usepackage{array}
\usepackage{rotating}
\usepackage{hyperref}
\usepackage{comment}
\usepackage{xspace}

\usepackage{xcolor}
\usepackage{listings}
\usepackage{longtable}
\usepackage{paralist}
\usepackage{lscape}
\usepackage{threeparttable}
\usepackage{supertabular}

\usepackage[tight,footnotesize]{subfigure}

\usepackage{pifont}% http://ctan.org/pkg/pifont
\usepackage{babel,adjustbox,booktabs,multirow}
\usepackage{lscape}
\usepackage{wasysym}
\usepackage{longtable}
\usepackage{multicol}
\usepackage{booktabs}
\usepackage{fontawesome} %\faCheck

\usepackage{algorithm} 
\usepackage[algo2e,ruled,vlined,linesnumbered]{algorithm2e}

\usepackage{soul} %for highlighting

\lstdefinestyle{searchstringstyle}{
	basicstyle=\ttfamily\scriptsize,
	breakatwhitespace=false,          		  breaklines=true,                 
	captionpos=b,                    
	numbers=none,                    
	numbersep=4pt,                  
	showspaces=false,                
	showstringspaces=false,
	showtabs=false,                  
	tabsize=2,
	frame=single
}

\newtcolorbox{shadedbox}{
	drop shadow southeast,
	breakable,
	enhanced jigsaw,
	colback=white,
	boxrule=0.80pt,
	left=0.3em,
	right=0.3em,
	top=0.1em,
	bottom=0.05em
}

\lstdefinestyle{PythonStyle} {
	backgroundcolor=\color{white},   
	commentstyle=\color{deepgreen}, 
	breakatwhitespace=false,
	keywordstyle=\color{deepblue},
	language=Python,
	stringstyle=\color{deepgreen},
	basicstyle=\footnotesize\ttfamily,
	showtabs=false,  
	tabsize=2,
	showstringspaces=false }

\newcommand*{\GV}{\texttt{Gavel}\@\xspace}
\newcommand*{\GA}{GHA\@\xspace}

\newcommand*{\GH}{GitHub\@\xspace}
\newcommand*{\RM}{\texttt{README.MD}\@\xspace}

\newcommand*{\CNB}{\texttt{CNBN}\@\xspace}

\newcommand*{\ie}{i.e.,\@\xspace}
\newcommand*{\eg}{e.g.,\@\xspace}
\newcommand*{\etal}{et al.\@\xspace}

\sethlcolor{lightgray} %changing the highlight color
\newcommand{\api}[1]{\texttt{\hl{\small #1}}}

\newcommand{\rqfirst}{\textbf{RQ$_1$}: \emph{Has the issue of categorizing \GH actions been investigated by Software Engineering research?}} 
\newcommand{\rqsecond}{\textbf{RQ$_2$}: \emph{Which configuration brings about the best prediction accuracy?}} 
\newcommand{\rqthird}{\textbf{RQ$_3$}: \emph{How does \GV compare with \CNB?}}

\definecolor{lightgray}{gray}{0.92}
\definecolor{verylightgray}{gray}{0.95}
\definecolor{deepblue}{rgb}{0,0,0.5}
\definecolor{deepred}{rgb}{0.6,0,0}
\definecolor{deepgreen}{rgb}{0,0.5,0}
\definecolor{shadecolor}{gray}{0.9}

\usepackage{pifont} 
\newcommand\untick{\ding{55}}

\begin{document} \sloppy

	\lstdefinestyle{JavaStyle} {
		backgroundcolor=\color{white},   
		commentstyle=\color{mygreen}, 
		breakatwhitespace=false,
		keywordstyle=\color{violet},
		language=Java,
		stringstyle=\color{blue},
		basicstyle=\scriptsize\ttfamily,
		%  frame=single,
		showstringspaces=false }

%%
%% The "title" command has an optional parameter,
%% allowing the author to define a "short title" to be used in page headers.

%\title{FILLER: A \SO Post Title Generator using Pre-Trained Language Models with Self Improvement and Post Ranking}

%\title{Do Pre-Trained Language Models Still Hold Potential in Generating Stack Overflow Post Titles?}

%\title{Boosting the Visibility of \GH Actions with Transformers and Few-shot Learning}

\title{Automatic Categorization of \GH Actions with Transformers and Few-shot Learning}

\author{Phuong T. Nguyen}

%\orcid{1234-5678-901}

\affiliation{%
	\institution{University of L'Aquila}
	% \streetaddress{P.O. Box 1212}
	\city{L'Aquila}
	%\state{Ohio}
	\country{Italy}
	%\postcode{43017-6221}
}
\email{phuong.nguyen@univaq.it}

\author{Juri Di Rocco}

%\orcid{1234-5678-901}

\affiliation{%
	\institution{University of L'Aquila}
	% \streetaddress{P.O. Box 1212}
	\city{L'Aquila}
	%\state{Ohio}
	\country{Italy}
	%\postcode{43017-6221}
}
\email{juri.dirocco@univaq.it}

\author{Claudio Di Sipio}

%\orcid{1234-5678-901}

\affiliation{%
	\institution{University of L'Aquila}
	% \streetaddress{P.O. Box 1212}
	\city{L'Aquila}
	%\state{Ohio}
	\country{Italy}
	%\postcode{43017-6221}
}
\email{claudio.disipio@univaq.it}

\author{Mudita Shakya}

%\orcid{1234-5678-901}

\affiliation{%
	\institution{University of L'Aquila}
	% \streetaddress{P.O. Box 1212}
	\city{L'Aquila}
	%\state{Ohio}
	\country{Italy}
	%\postcode{43017-6221}
}
\email{mudita.shakya@student.univaq.it}

\author{Davide Di Ruscio}
%\orcid{1234-5678-901}
\affiliation{%
	\institution{University of L'Aquila}
	% \streetaddress{P.O. Box 1212}
	\city{L'Aquila}
	%\state{Ohio}
	\country{Italy}
	%\postcode{43017-6221}
}
\email{davide.diruscio@univaq.it}

\author{Massimiliano Di Penta}
\affiliation{%
	\institution{University of Sannio}
%	\institution{Universit\`a degli studi del Sannio}
	\city{Benevento}
	\country{Italy}
}
\email{dipenta@unisannio.it}

\renewcommand{\shortauthors}{ et al.}

%%
%% The abstract is a short summary of the work to be presented in the
%% article.
\begin{abstract}
In the \GH ecosystem, workflows are used as an effective means to automate development tasks and to set up a Continuous Integration and Delivery (CI/CD pipeline).
\texttt{\GH Actions} (\GA) have been conceived to
provide developers with a practical tool to create and maintain workflows, avoiding ``reinventing the wheel'' and cluttering the workflow with shell commands.
Properly leveraging the power of \texttt{\GH Actions} can facilitate the development processes, enhance collaboration, and significantly impact project outcomes. To expose actions to search engines, \GH allows developers to assign them to one or more categories manually. These 
are used as an effective means to group actions sharing similar functionality. 
Nevertheless, while providing a practical way to execute workflows, many actions have unclear purposes, and sometimes they are not categorized.	
In this work, we bridge such a gap by conceptualizing \GV, a practical solution to increasing the visibility of actions in \GH. By leveraging the content of \RM files for each action, we use Transformer--a deep learning algorithm--to assign suitable categories to the action. We conducted an empirical investigation and compared \GV with a state-of-the-art baseline. The experimental results show that our proposed approach can assign categories to \GH actions effectively, thus outperforming the state-of-the-art baseline.
\end{abstract}
%%
%% The code below is generated by the tool at http://dl.acm.org/ccs.cfm.
%% Please copy and paste the code instead of the example below.
%%
%\begin{CCSXML}
%<ccs2012>
% <concept>
%  <concept_id>00000000.0000000.0000000</concept_id>
%  <concept_desc>Do Not Use This Code, Generate the Correct Terms for %Your Paper</concept_desc>
%  <concept_significance>500</concept_significance>
% </concept>
% <concept>
%  <concept_id>00000000.00000000.00000000</concept_id>
%  <concept_desc>Do Not Use This Code, Generate the Correct Terms for %Your Paper</concept_desc>
%  <concept_significance>300</concept_significance>
% </concept>
% <concept>
%  <concept_id>00000000.00000000.00000000</concept_id>
%  <concept_desc>Do Not Use This Code, Generate the Correct Terms for %Your Paper</concept_desc>
%  <concept_significance>100</concept_significance>
% </concept>
% <concept>
%  <concept_id>00000000.00000000.00000000</concept_id>
%  <concept_desc>Do Not Use This Code, Generate the Correct Terms for %Your Paper</concept_desc>
%  <concept_significance>100</concept_significance>
% </concept>
%</ccs2012>
%\end{CCSXML}

%\ccsdesc[500]{Do Not Use This Code~Generate the Correct Terms for Your %Paper}
%\ccsdesc[300]{Do Not Use This Code~Generate the Correct Terms for Your %Paper}
%\ccsdesc{Do Not Use This Code~Generate the Correct Terms for Your Paper}
%\ccsdesc[100]{Do Not Use This Code~Generate the Correct Terms for Your %Paper}

%%
%% Keywords. The author(s) should pick words that accurately describe
%% the work being presented. Separate the keywords with commas.

% \received{20 February 2007}
% \received[revised]{12 March 2009}
% \received[accepted]{5 June 2009}

%%
%% This command processes the author and affiliation and title
%% information and builds the first part of the formatted document.
\maketitle

\section{Introduction}
\label{sec:introduction}

%==================================
%Read more about this: https://machinelearningmastery.com/one-vs-rest-and-one-vs-one-for-multi-class-classification/
%==================================

Within a \GH repository, a workflow consists of a series of automated steps executed in response to specific events. Workflows can be initiated through triggering events, including code commits, issues, pull requests, comments, or scheduled events, to name a few \cite{10.1145/3597503.3623351}. This facilitates
the automation of various aspects of software development and project management. In fact, workflows simplify the automation of specific tasks (\eg configuring a particular 
environment, releasing software on a package registry, running tests, or assessing code quality) without the need to write custom code for each task.
In \GH, workflows are the premier way to automate CI/CD (Continuous Integration and Delivery).
Workflows steps are defined using \texttt{\GH Actions} 
or other tools invoked as commands~\cite{10.1145/3593434.3593475}. 
Actions are reusable assets that can be used in place of custom scripts and are available in public \GH repositories and a specific \GH 
Marketplace.\footnote{\url{https://github.com/marketplace/}} %This innovative addition can have profound implications for organizations and project management. 

\begin{figure*}[t!]
	\centering
	\begin{tabular}{c c }	
		\subfigure[An action with categories]{\label{fig:FullDetails}
			\includegraphics[width=0.40\linewidth]{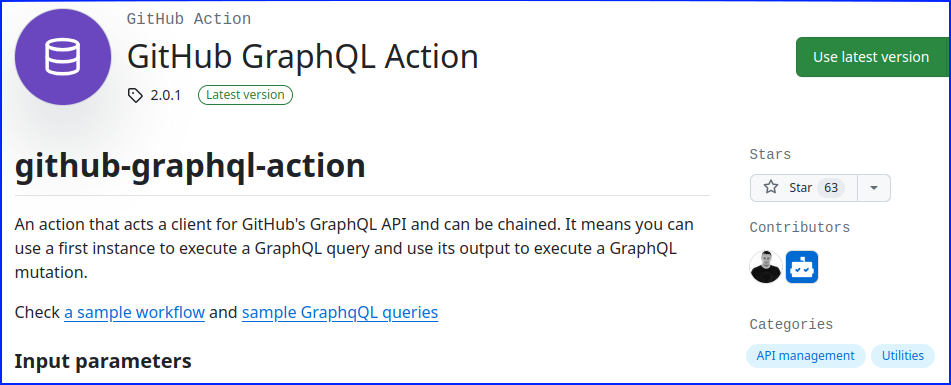}}	&
		\subfigure[An action with no categories]{\label{fig:NoDetails}\includegraphics[width=0.40\linewidth]{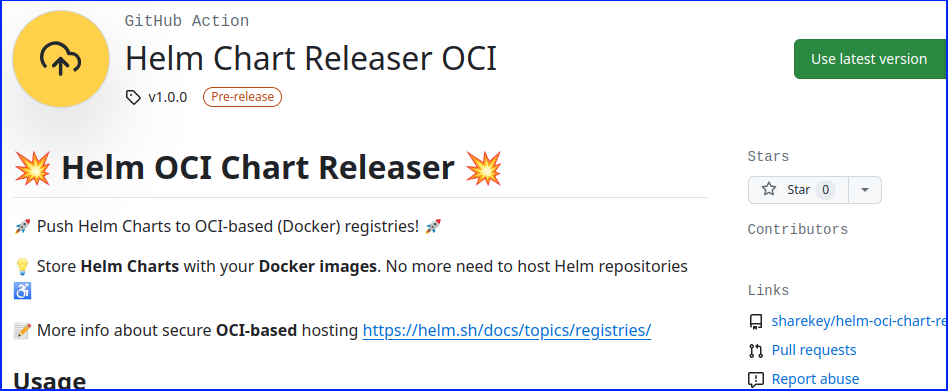}} 
	\end{tabular} 
	\vspace{-.2cm}
	\caption{Examples of \GH actions with and without categories.} 	
	\label{fig:MotivatingExample}
	\vspace{-.4cm}
\end{figure*}

When uploading an action to the Marketplace, developers usually assign it 
to some \textit{categories} to characterize its constituent functionality.  
Essentially, these categories are an effective means to group similar actions into common classes, allowing other developers to approach the ones that match their needs. This fosters reusability among users in the open-source community~\cite{10.1145/3661167.3661215}, providing them with artefacts relevant to the considered tasks. Nevertheless, while offering a practical way to ease the development activities, many actions are not visible to users, \ie 
several of them 
are not specified to any categories, being 
obscure to search engines. 
Therefore, it would be important to support developers in effectively defining categories for the \GH actions they publish on the Marketplace or on \GH in general.

By means of a lightweight systematic literature review %~\cite{KITCHENHAM20097} 
(SLR), in this work we found out that no Software Engineering (SE) work has ever been conducted to automatically attach categories to \GH actions. Being motivated by this gap, we conceive \GV,\footnote{Gavel is a hammer usually made of wood, and used in events like a court, or an auction to attract attention from the audience.} 
a \underline{G}itHub \underline{a}ctions \underline{v}isibility \underline{el}evator based on transformers and few-shot learning to make actions more visible and accessible 
to developers. Given an action, we use its \RM as input data, and build a multi-label classifier combined with few-shot learning 
to assign one or more categories to it. 
We evaluate the performance of \GV 
using a dataset collected from \GH, and compare it with a quasi-equivalent tool. The experiments show that \GV is effective at assigning categories to actions, obtaining an encouraging prediction, thereby outperforming the baseline. 

The contributions of our work are summarized as follows:

\begin{itemize}
	\item An approach named \GV built on top of sentence transformers to equip \GH actions with proper categories to increase their visibility.
	\item An empirical evaluation of \GV on real-world datasets, and comparison with a state-of-the-art baseline.
	\item The tool together with the data curated in this work is published online to enable future research~\cite{GavelReplicationPackage}.
\end{itemize}

\begin{comment}
\noindent
\textbf{Structure.} Section~\ref{sec:Background} introduces a motivating example together with the related background. The proposed approach is described in Section~\ref{sec:ProposedApproach}, while  
Section~\ref{sec:Evaluation} presents the materials and methods used to conduct an empirical evaluation on the proposed approach. Afterward, Section~\ref{sec:Results} reports and analyzes the
results. 
Finally, Section~\ref{sec:Conclusion} concludes the paper and outlines 
future work.
\end{comment}

\section{Motivation and background}
\label{sec:Background}

In this section, we take a concrete example to highlight %illustrate 
the importance %necessity 
of categories for \GH actions. Afterward, we review the related studies as a base for further presentation. %to position our work.

\subsection{Motivating example} \label{sec:MotivatingExample}

%\begin{figure}[!h]
%	\centering
%%	\vspace{-.2cm}
%	\includegraphics[width=0.48\textwidth]{figs/MotivatingExample.png}
%	\caption{An action with full details.}
%	\label{fig:MotivatingExample1}
%%	\vspace{-.4cm}
%\end{figure}
%
%
%\begin{figure}[!h]
%	\centering
%%	\vspace{-.2cm}
%	\includegraphics[width=0.48\textwidth]{figs/MotivatingExample2.png}
%	\caption{An action without categories and description.}
%	\label{fig:MotivatingExample2}
%%	\vspace{-.4cm}
%\end{figure}

Figure~\ref{fig:MotivatingExample} displays 
a motivating example with 2 actions. The first one named \texttt{GitHub GraphQL Action}\footnote{\url{https://github.com/marketplace/actions/github-graphql-action}}  
(see Figure~\ref{fig:FullDetails}) is employed in executing a GraphQL query, and using its output to run a GraphQL mutation.
The action has been classified into two categories, \ie \texttt{API management} and \texttt{Utilities}. When we expand one of the categories, we will be redirected to a set of all the actions belonging to the same category. This means that whenever users look into either the \texttt{API management} or the \texttt{Utilities} category, they will see \texttt{GitHub GraphQL Action} among the entries.
The action in Figure~\ref{fig:NoDetails} with the name \texttt{Helm Chart Releaser OCI}\footnote{\url{https://github.com/marketplace/actions/helm-chart-releaser-oci}} can be used to push Helm Chart to container registries, including those hosted in \GH or Azure. Noteworthy, this action has not been sorted into any categories, and due to this, it will never show up in the search towards the available categories in Marketplace. Essentially, this limits the action visibility and, consequently, its adoption.

%Through the %this motivating 
%example, we see 

The example implies 
the need for a proper technique to recommend categories for actions.  
%that have not been provided with any labels. 
We conjecture that 
the \RM file(s) of an action can be used as input for a recommender 
%process 
%to generate %relevant 
of categories. This requires suitable categorization algorithms  
able to perform 
multi-label classification. %In the next subsection, we review the related work that deals with the classification of various artifacts in \GH. 

\subsection{Related Work} \label{sec:RelatedWork}

%====================================
%In this section, we review notable studies that are relevant to ours. First, we present approaches that perform multi-class classification using \RM and additional textual content. Afterward, approaches that support the automation of \GH workflows are reviewed. 
%====================================

%====================================
%A qualitative study involving the manual annotation of the content of 4,226 sections from 393 randomly selected GitHub README files, establishing a point of reference for the content of a GitHub README file. We distinguish eight categories in the coding schema that emerged from our qualitative analysis (What, Why, How, When, Who, References, Contribution, and Other), and we report their respective frequencies and associations.
%====================================

%\paragraph{Categorization of \RM files} %\label{sec:readme}

%The work of David Lo.

Prana \etal~\cite{DBLP:journals/ese/PranaTTAL19} conducted a qualitative study to manually label the content of 
393 randomly selected \RM files. %, creating a point of reference for the content of a \RM. 
They identified eight	categories in the coding schema coming from a qualitative analysis, %\ie \emph{What}, \emph{Why}, \emph{How}, \emph{When}, \emph{Who}, \emph{References}, \emph{Contribution}, and \emph{Other}, 
and reported the respective frequencies and associations. %Moreover, 
They also developed an approach to the categorization of \RM files. GHTRec~\cite{9590294} predicts the list of topics using the preprocessed 
content using the BERT model. Furthermore, the tool suggests the most similar trending repositories, namely the most starred ones, by computing the similarity on the topic vectors mined from the user's historical data. 

Izadi \etal~\cite{izadi2020topic} leveraged a fine-tuned version of BERT~\cite{bert2018} to classify \GH repositories using textual content.
The training data includes 152K GitHub repositories and 228 featured topics. Then, the tool exploits a fine-tuned version of DistilBERT to classify a 
repository from all the collected textual data, \ie \RM content, project name, descriptions, wiki pages, and file names. The same authors enriched the 
classifier by exploiting semantic relationships \cite{izadi_semantically-enhanced_2023}. To this end, a knowledge graph composed of 2,234 semantic relationships and 863 curated topics was built according to \textit{(i)} their popularity and \textit{(ii)} the degree in the produced graph. The novel approach, called SED-KG, is 
employed to enhance the performance of existing topic recommender systems. 

HybridRec \cite{di_rocco_hybridrec_2022} predicts a list of topics for a given \GH
repository by relying on 2 components, \ie a stochastic classifier and a collaborative filtering engine. The former is a Complement Na\"ive Bayesian  network (\CNB) to encode \RM content using TF-IDF, 
predicting an initial set of topics. The latter exploits such predictions to infer additional tags, thus increasing the coverage by considering the mutual relationships encoded as a graph. 

%======================
%ClassifyHub~\cite{soll2017classifyhub} is an ensemble learning approach that relies on the InformatiCup 2017 dataset\footnote{\href{https://github.com/informatiCup/informatiCup2017}{https://github.com/informatiCup/informatiCup2017}} to perform the text classification task.  
%======================

GitRanking \cite{sas_gitranking_2023} uses the active sampling algorithm for Pairwise comparisons (ASAP) to build a topic taxonomy by combining \RM files and Wikidata definition. %After the data collection phase, 
A manual filtering is applied by involving real developers in annotating the most 3,000 popular topics. Then, ASAP is used to perform the final ranking by leveraging the WikiData description of each topic. %The result of this is an online framework for creating software categorization. 

Mastropaolo \etal \cite{10.1145/3597503.3623351} presented GH-WCOM (GitHub Workflow COMpletion), a Transformer-based approach to assist developers in specifying 
\GH workflows. The authors also conducted a qualitative analysis to investigate the extent to which the recommendations provided by GH-WCOM are useful when the generated output is different from the target one. We conjecture that \GV can be complementary to GH-WCOM, and possibly improve its action suggestions.

By reviewing the related studies, 
we see that while %being 
relevant to our work in terms of tasks, the aforementioned pieces of work deal with the classification of other types of artifacts---such as \GH repositories---rather than \GH actions, %
%while they obtain encouraging performance, 
and none of them has tackled %dealt with 
the lack of training data. %Moreover,
Among others, only \CNB~\cite{di_rocco_hybridrec_2022} and Repository Catalogue~\cite{izadi_semantically-enhanced_2023,izadi2020topic} work with multi-label classification of artifacts, albeit not \GH actions. 
%By means of 
Through an evaluation, we found out that they treat a \RM file as a whole, without distinguishing different elements such as code, or text, 
failing to categorize \GH 
actions. 
%Altogether, 
This triggers the need for a tool to assign 
labels to actions, that should be able to handle diverse components of \RM files, as well as to rectify the lack of labeled data.

%\MAX{reviewer 2 can still say the problem is similar to repository classification. we should try to make the case that it is different}

%\paragraph{Specification of \GH workflows by means of actions} %\label{sec:workflows}
%
%
%Mastropaolo \etal \cite{10.1145/3597503.3623351} presented GH-WCOM (GitHub Workflow COMpletion), a Transformer-based approach supporting developers in writing a specific type of CI/CD pipelines, namely GitHub workflows. 
%
%This work~\cite{10.1145/3593434.3593475}.

\section{Proposed Approach}
\label{sec:ProposedApproach}

%To bridge the aforementioned gap, 
%To tackle the 
%In this paper, we develop 

This section describes in detail \GV--our conceived tool for increasing the visibility of \GH actions built on top of transformers and few-shot learning. Given an action, using the \RM file(s) as the sole input, \GV generates 
a list of possible categories to which the action should belong.

\begin{figure}[h!]
	\centering
	\vspace{-.1cm}
	\includegraphics[width=0.48\textwidth]{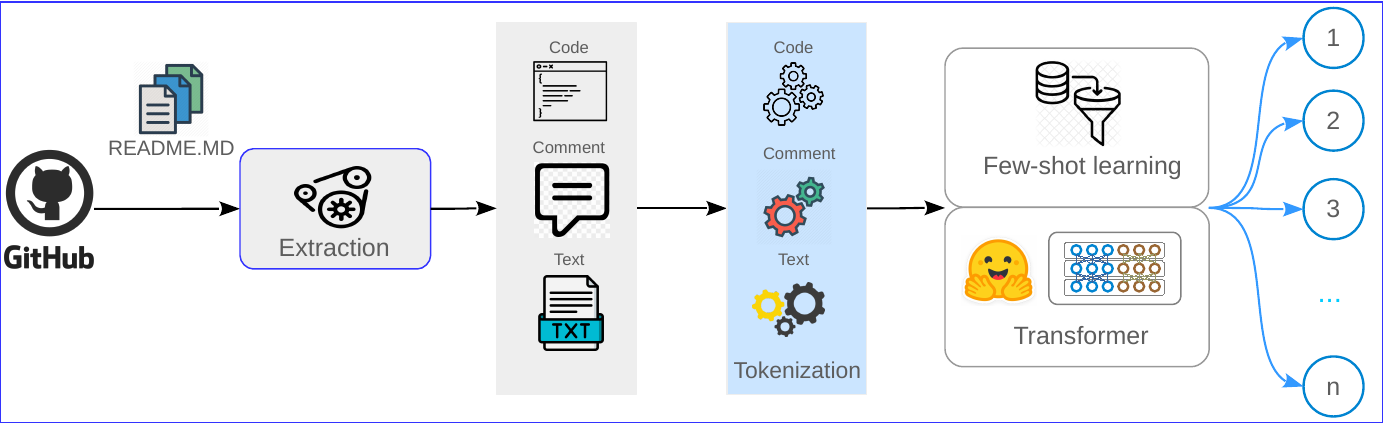}
	\caption{System architecture.}
	\vspace{-.2cm} 
	\label{fig:System}
	%	\vspace{-.4cm} 
\end{figure}

Figure~\ref{fig:System} depicts the proposed architecture with four main modules: Extraction, Tokenization, Classification, and Few-shot Learning, 
explained in the following subsections.

\subsection{Extraction and Tokenization}

In Figure~\ref{fig:System}, the first phase involves parsing data from raw \RM files. 
To extract the needed data to separate fields, \ie code, comments, and text from files for the classification task, we used a dedicated Python library for \textit{.MD} files, named Marko.\footnote{\url{https://marko-py.readthedocs.io/en/latest/index.html}} Compared to existing libraries \cite{noauthor_python-markdown_2023,yang_mistune_2023}, it is compliant with the latest \textit{CommonMark} specifications,\footnote{\url{https://spec.commonmark.org/0.30/}} thus improving the overall parsing. 

The resulting data is then encoded using a tokenizer,
generating a numerical representation that captures the meaning of the input. 
Tokenization is a crucial pre-processing step to prepare %sequential 
data for DL 
models. By breaking down the data into 
the constituent parts, containing words or sub-words called tokens, the tokenization enables models to learn from the structure and semantics of the data, leading to better performance for various tasks~\cite{FU2024122563}. 

\begin{lstlisting}[caption={Example of code and comments.},style=PythonStyle, captionpos=t, label={lst:OriginalCode}, breaklines=true, language=python, xleftmargin=0em,frame=single, escapechar=\|,numbers=none,backgroundcolor=\color{verylightgray}]
	# This function greets the user
	def greet(name):
	print(f"Hello, {name}!")
	greet("Alice")
\end{lstlisting}

%\begin{lstlisting}[caption=The transformed sequence.,basicstyle=\footnotesize\ttfamily,label={lst:TransSequence},escapechar=£,backgroundcolor=\color{verylightgray},frame=single,tabsize=2]
%	BOS, # This function greets the user def greet(name):
%	print(f"Hello, {name}!") greet("Alice"), EOS
%\end{lstlisting}

%As an example to show how tokenization works,
 Listing~\ref{lst:OriginalCode} depicts a simple code snippet, together with code comments. The %corresponding 
 transformed sequences for comments and code are: % as follows: 
 \api{BOS,\; \#,\; This,\; function,\; greets,\; the,\; user,\; EOS} and \api{BOS,\; def,\; greet,\; (,\; name,\; ),\; :,\; print,\; (,\; f,\; ",\; Hello,\; ,\; {name}! "),\; greet,\; (",\; Alice,\; "),\; EOS}. 
BOS and EOS are two special delimiters %used 
to signify the beginning and end of a sequence. %	, respectively.

%\begin{figure}[h!]
%	\centering
%	\vspace{-.1cm}
%	\includegraphics[width=0.35\textwidth]{figs/CodeExample.pdf}
%	\caption{Example of code.}
%%	\vspace{-.2cm} 
%	\label{fig:CodeExample}
%%		\vspace{-.4cm} 
%\end{figure}

\subsection{Multi-label Classification}

Once the data has been collected and encoded with the tokenizer, it is necessary to parse them with a suitable format. 
The input contains $X$ as a set of $m$ \RM files, together with matrix $Y[m,n]$ to store the labels, \ie categories. Each row specifies the categories of a repository: if the repository belongs to a category, then the corresponding column is set to 1, otherwise it is set to 0.
The pseudo-code in Algorithm~\ref{alg:OneVsOneClassification} illustrates how the process unfolds.
Starting from these matrices, we train a set of classifiers, adopting the one vs. rest classification paradigm~\cite{PAWARA2020107528}.  For each class, a classifier assigns each sample to one or more categories, which 
are not mutually exclusive. Due to space limit, we cannot recall the algorithm here. Interested readers are kindly referred to the work of Pawara et al.~\cite{PAWARA2020107528} for details.

\begin{algorithm2e}[t!]
	\caption{One vs. all multi-label classification} %
	\label{alg:OneVsOneClassification}
	\DontPrintSemicolon
	\SetAlgoLined
	%\KwResult{Write here the result}
	\SetKwInOut{Input}{Input} 
	\SetKwInOut{Output}{Output}
	\Input{
		\begin{itemize}
			\item X: $m$ \RM files containing text, code, comments. %The fine--tuned model $\mathcal{M}_{\theta_f}$
			\item Y: Matrix of categories of size [$m$, $n$]. % of the \RM files. %The original training dataset $\mathcal{D}_{t}$
			%			\item $\mathcal{C}$: a classifier.
		\end{itemize}
	}
	\Output{
		\begin{itemize}
			\item $\mathcal{C} = [c_1..c_n]$: A set of $n$ binary classifiers, each corresponds to one category.
			%			\item The new fine-tuned model $\mathcal{M}_{\theta_{f^*}}$
		\end{itemize}
	}
	$\mathcal{C} \leftarrow \emptyset$
	
	\ForEach {$i \in [1..n]$}
	{
		%		Initialize an empty list classifiers
		$c_i \leftarrow \emptyset$
		
		$y_i \leftarrow Y[:, i]$
		
		$c_i \leftarrow Train(X, y_i)$ %using X and $y_i$
		
		$\mathcal{C} \leftarrow \mathcal{C}  \bigcup c_i$
		
		%		Feature matrix of shape (n\_samples, n\_features)
		
		%		Label matrix of shape (n\_samples, n\_labels)
	}
	\Return{$\mathcal{C}$}
\end{algorithm2e}

\vspace{-.2cm}

\subsection{Few-shot learning}

Deep learning models usually require a large amount of data for training~\cite{DBLP:journals/sncs/Sarker21c}. To feed supervised learning algorithms, data needs to be properly labeled, which is usually done by humans. In fact, collecting data is a error-prone and daunting task. To this end, few-shot learning has been proposed as a practical solution to the lack of training data~\cite{BLAES2017159}, allowing a model to be trained with a small amount of data. For each class, only a few samples are selected to train the model, and the rest is used for fine tuning. This simulates the case where only a limited amount of data is available. %\MAX{It's not very clear how the few shot was performed, it's not explained}
Such a model aims to learn better by using relevant, and well-curated samples, rather than low-quality data~\cite{DBLP:journals/mta/YanZC18}. Moreover, training or fine-tuning with less data helps save time and resources. 

For the implementation of \GV, we make use of \api{SetFit},\footnote{\url{https://huggingface.co/docs/setfit/index}} a pre-trained transformer model of text hosted in Hugging Face. 
%\MAX{is SetFit an API, a model, or both?} \MAX{pretrained on what?} 
Since the fine-tuning is resource demanding, 
\GV was trained and tested on a Linux High-Performance Computing (HPC) cluster. Instead, \CNB\footnote{We use the original implementation made available by the authors at: \url{https://github.com/MDEGroup/HybridRec}} is more lightweight, and it was run on a local machine with OS Windows 11 Processor AMD Ryzen 9 RAM 16 GB.

\section{Evaluation Methodology} %Evaluation materials and methods}
\label{sec:Evaluation}

This section describes in detail the materials and methods used to evaluate 
\GV, and compare it with a state-of-the-art baseline.

\subsection{Research Questions}

\smallskip
\noindent
$\triangleright$ \rqfirst~In this research question, we perform a lightweight systematic literature review to investigate to what extent the issue of categorization of \GH actions has been studied by state-of-the-art research.

\smallskip
\noindent
$\triangleright$ \rqsecond~\GV can accept as input different types of \RM data, including text, code, or code together with comments, or all of them. We empirically evaluate which of these configurations is more effective, bringing a superior performance with respect to prediction accuracy. 

\smallskip
\noindent
$\triangleright$ \rqthird~To the best of our knowledge, no work has ever been conducted to categorizing \GH actions, thus for comparison, in this work we have to consider \CNB~\cite{di_rocco_hybridrec_2022}, a baseline that has been conceived for a different domain, \ie categorization of \GH topics.

\subsection{Metrics and Datasets}

\smallskip
\noindent
$\triangleright$ \emph{Metrics.} There are the following definitions: %The metrics used in this paper are based on: % the following numbers: 
%We define the 
True positives (TP): the number of correctly predicted categories; False positives (FP): the number of incorrectly predicted categories; False negatives: the number of incorrectly not predicted categories. Then, the metrics widely used to assess classification approaches are defined as follows \cite{buckland_relationship_1994}: Precision $ P = TP / (TP + FP) $; Recall $ R = TP / (TP + FN)$; and F1-score $ F1 = 2* P * R /(P + R)$. 

%====================================
%\begin{multicols}{3}
%	\noindent
%	\begin{align}\nonumber
	%		\text{P} &= \frac{\text{TP}}{\text{TP} + \text{FP}} 
	%%		\label{eq:precision}
	%	\end{align}
%	
%	\noindent
%	\begin{align}\nonumber
	%		\text{R} &= \frac{\text{TP}}{\text{TP} + \text{FN}} 
	%%		\label{eq:recall}
	%	\end{align}
%	
%	\noindent
%	\begin{align}\nonumber
	%		\text{F1} &= 2 \times \frac{\text{P} \times \text{R}}{\text{P} + \text{R}}
	%%		\label{eq:f1}
	%	\end{align}
%\end{multicols}
%\vspace{-.4cm}
%====================================

For the presentation, we use the micro average, macro average, and weighted average score of these metrics. Micro average assigns an equal weight to each instance, irrespective of the category and the number of samples. 
Macro average is the arithmetic mean of all the scores for the categories, and weighted average gauges the varying degree of importance of the categories in a dataset.

%\subsubsection*{Estimating time performance.}
% Besides the accuracy evaluation, we assess the time for the training and testing phases of the two considered models, \ie the CNB and \GV. Since the fine-tuning is demanding from the computational point of view, we employ a Linux High-Performance Computing (HPC) cluster. Instead, we recorded the time required by the CNB on a local machine with Windows 11 OS, AMD Ryzen 9 Processor, and 16 GB RAM.

%Laboratory of the Department of Information Engineering, Computer
%Science and Mathematics (DISIM) at the University of L’Aquila

%To the best of our knowledge, there has been no approach conceived to the categorization of \GH actions, thus we cannot find any baselines in the same domain.

\smallskip
\noindent
$\triangleright$ \emph{Dataset.} As \GH does not provide an API 
to retrieve actions from Marketplace, we had to collect data from: \emph{(i)} %different sources. 
a dataset~\cite{DECAN2023111827} %we use 
consists of 958 actions; and \emph{(ii)} the second one~\cite{9463074} includes 708 actions. 
We combined them to create a single dataset. 
This is done by 
aligning the actions' metadata, resolving discrepancies, and removing duplicates. 
The final dataset contains 1,213 unique actions, 
spanning in 
30 categories, and covering diverse 
functionality.

\subsection{Configurations}

The \RM file(s) of an action may contain different types of data, including text, code, and comments. Moreover, sometimes they also feature media files %, \eg audio, or video, 
to provide a visualization of how to set up or execute the workflows. In %the scope of 
this paper, we consider the following regular  
data types: \emph{(i)} \emph{Text}. Most \RM files contain text to describe the main functionalities of actions; \emph{(ii)} \emph{Source code}. \RM files usually include source code as a means to illustrate how a workflow can be executed; and \emph{(iii)} \emph{Code comments}. They are embedded text to describe the corresponding source code.

%To extract the needed data for the classification task, we used a dedicated Python library for \textit{.md} files, named Marko.\footnote{\url{https://marko-py.readthedocs.io/en/latest/index.html}} Compared to existing libraries \cite{noauthor_python-markdown_2023,yang_mistune_2023}, it is compliant with the latest \textit{CommonMark} specifications,\footnote{\url{https://spec.commonmark.org/0.30/}} thus improving the overall parsing. 
%We filtered out special characters, such as line breaks or empty lines. Thus, a \textit{cell} is a row of the \RM with actual content. In the scope of the work, we limited ourselves to two main types of cells, \ie code and text cells, as depicted in Figure \ref{fig:data_model}. In particular, we could extract 526,685 cells classified as code and 842,628 cells containing textual content. The \RM miner has been used to extract three different configurations for all datasets, \ie code, text, and global. 

%, including the following ones: %of input data in our experiments. There are the following elements:

%%\begin{itemize}
%\smallskip
%\noindent
%$\triangleright$ \emph{Text}. Most \RM files contain text to describe the main functionalities of actions.
%
%\smallskip
%\noindent
%$\triangleright$ \emph{Source code}. \RM files usually include source code as a means to illustrate how a workflow can be executed.
%
%\smallskip
%\noindent
%$\triangleright$ \emph{Code comments}. They are embedded text to give a description for the corresponding source code.
%%\end{itemize}

\begin{table}[h!]
	\centering
	\scriptsize	
	\vspace{-.2cm}
	\caption{Experimental configurations.}
	\vspace{-.2cm}
	\begin{tabular}{|p{1.00cm} | p{1.20cm} | p{1.20cm} | p{1.20cm} | p{1.60cm} |}	\hline
		%		& \multicolumn{3}{c|}{\textbf{Dataset}} \\ \cline{2-4}
		%		& \textbf{DS$_1$}~\cite{NGUYEN2019110460}  & \textbf{DS$_2$}  & \textbf{DS$_3$}~\cite{9043686,10.1145/3468264.3468552}  \\ \hline %\hline
		%		\# projects & 1,200 & 5,200 & 56,091 \\ %\hline
		%		\# libraries & 13,497 & 31,817 & 762 \\ %\hline
		%		Type & Generic software & Generic software & Android apps \\ \hline %\hline
		%		\textbf{System} & \multicolumn{3}{c|}{\textbf{Experimental configurations}} \\ \hline
		\rowcolor{lightgray} Conf. & Text & Code & Comments & \# of samples \\ \hline
		C$_1$ & \faCheck & \faCheck & \faCheck & 1,213\\ \hline
		C$_2$ & \untick & \faCheck & \faCheck & 1,140 \\ \hline
		C$_3$ & \untick & \untick & \faCheck & 436 \\ \hline
		C$_4$ & \untick & \faCheck & \untick & 1,140 \\ \hline		
		C$_5$ & \faCheck & \untick & \untick & 1,210 \\ \hline
	\end{tabular}
	\vspace{-.2cm}
	\label{tab:Configurations}
\end{table}

Table~\ref{tab:Configurations} lists the 
configurations considered in our experiments. 
By C$_1$ all data, including code, comments, and text in \RM is used as input, while with C$_2$, text is not considered. By C$_3$, comments
in code are extracted and used as the sole input. Code is the only artifact considered as input by C$_4$. Eventually, in C$_5$, only textual data in \RM is parsed and fed to train \GV. Depending on the availability of data, many 
files do not include some of the aforementioned artifacts. This is the reason why the number of samples--depicted in the last column of the table--varies by the configurations. For instance, there are only 436 samples for C$_3$ as few \RM files have comments embedded in code.

\section{Results and Discussion}
\label{sec:Results}
\begin{table*}[h!]
\caption{Classification report for C$_1$, C$_2$, C$_3$, and C$_4$.}
\label{table:classification_all}
%\scriptsize
\footnotesize
\centering
\begin{tabular}{|l | c| c | c | r | c| c | c | r | c| c | c | r | c| c | c | r |} \hline
	 & \multicolumn{4}{c|}{\textbf{C$_1$}}  & \multicolumn{4}{c|}{\textbf{C$_2$}} & \multicolumn{4}{c|}{\textbf{C$_3$}} & \multicolumn{4}{c|}{\textbf{C$_4$}} \\ \hline
	\textbf{Category} & P & R & F1 & \# & P & R & F1 & \#  & P & R & F1 & \#  & P & R & F1 & \# \\ \hline
%	\midrule
	\rowcolor{lightgray}
	AI Assisted & 1.00 & 0.20 & 0.33 & 5 & 1.00 & 0.20 & 0.33 & 5 & 1.00 & 1.00 & 1.00 & 1 & 1.00 & 0.20 & 0.33 & 5\\
	API management & 1.00 & 0.07 & 0.13 & 14 & 0.50 & 0.07 & 0.12 & 14 & 1.00 & 1.00 & 1.00 & 1 & 1.00 & 0.07 & 0.13 & 14\\
	\rowcolor{lightgray}
	Backup Utilities & 1.00 & 0.50 & 0.67 & 2 & 1.00 & 0.50 & 0.67 & 2 & 1.00 & 1.00 & 1.00 & 1 & 1.00 & 0.50 & 0.67 & 2\\
	Chat & 0.72 & 0.57 & 0.64 & 37 & 0.84 & 0.57 & 0.68 & 37 & 1.00 & 1.00 & 1.00 & 1 & 0.83 & 0.65 & 0.73 & 37\\
	\rowcolor{lightgray}
	Code Scanning Ready & 1.00 & 0.33 & 0.50 & 3 & 1.00 & 0.33 & 0.50 & 3 & 1.00 & 1.00 & 1.00 & 1 & 1.00 & 0.33 & 0.50 & 3\\
	Code quality & 0.76 & 0.84 & 0.80 & 129 & 0.75 & 0.81 & 0.78 & 123 & 0.44 & 0.88 & 0.58 & 8 & 0.76 & 0.81 & 0.78 & 123\\
	\rowcolor{lightgray}
	Code review & 0.75 & 0.67 & 0.71 & 103 & 0.71 & 0.65 & 0.68 & 97 & 0.50 & 0.25 & 0.33 & 8 & 0.74 & 0.69 & 0.71 & 97\\
	Code search & 1.00 & 0.12 & 0.22 & 8 & 1.00 & 0.17 & 0.29 & 6 & 1.00 & 1.00 & 1.00 & 1 & 1.00 & 0.17 & 0.29 & 6\\
	\rowcolor{lightgray}
	Community & 1.00 & 0.08 & 0.15 & 25 & 0.50 & 0.05 & 0.08 & 22 & 1.00 & 1.00 & 1.00 & 1 & 1.00 & 0.09 & 0.17 & 22\\
	Container CI & 0.56 & 0.15 & 0.24 & 60 & 0.50 & 0.09 & 0.15 & 56 & 1.00 & 0.25 & 0.40 & 4 & 0.59 & 0.18 & 0.27 & 56\\
	\rowcolor{lightgray}
	Continuous integration & 0.86 & 0.95 & 0.90 & 434 & 0.89 & 0.94 & 0.91 & 407 & 0.44 & 0.67 & 0.53 & 30 & 0.87 & 0.95 & 0.91 & 407\\
	Dependency management & 0.88 & 0.65 & 0.75 & 89 & 0.75 & 0.57 & 0.65 & 82 & 0.10 & 0.33 & 0.15 & 3 & 0.73 & 0.59 & 0.65 & 82\\
	\rowcolor{lightgray}
	Deployment & 0.75 & 0.84 & 0.79 & 194 & 0.77 & 0.81 & 0.79 & 189 & 0.70 & 0.33 & 0.45 & 21 & 0.75 & 0.84 & 0.79 & 189\\
	Desktop tools & 1.00 & 0.10 & 0.18 & 10 & 1.00 & 0.10 & 0.18 & 10 & 1.00 & 1.00 & 1.00 & 1 & 1.00 & 0.10 & 0.18 & 10\\
	\rowcolor{lightgray}
	GitHub Sponsors & 1.00 & 0.50 & 0.67 & 2 & 1.00 & 0.50 & 0.67 & 2 & 1.00 & 1.00 & 1.00 & 1 & 1.00 & 0.50 & 0.67 & 2\\
	IDEs & 1.00 & 0.17 & 0.29 & 6 & 1.00 & 0.17 & 0.29 & 6 & 1.00 & 1.00 & 1.00 & 1 & 1.00 & 0.17 & 0.29 & 6\\
	\rowcolor{lightgray}
	Learning & 1.00 & 0.33 & 0.50 & 3 & 1.00 & 0.33 & 0.50 & 3 & 1.00 & 1.00 & 1.00 & 1 & 1.00 & 0.33 & 0.50 & 3\\
	Localization & 1.00 & 0.25 & 0.40 & 4 & 1.00 & 0.33 & 0.50 & 3 & 1.00 & 0.50 & 0.67 & 2 & 1.00 & 0.33 & 0.50 & 3\\
	\rowcolor{lightgray}
	Mobile & 1.00 & 0.20 & 0.33 & 5 & 1.00 & 0.20 & 0.33 & 5 & 1.00 & 1.00 & 1.00 & 1 & 1.00 & 0.20 & 0.33 & 5\\
	Mobile CI & 0.67 & 0.22 & 0.33 & 9 & 1.00 & 0.25 & 0.40 & 8 & 1.00 & 1.00 & 1.00 & 1 & 1.00 & 0.12 & 0.22 & 8\\
	\rowcolor{lightgray}
	Monitoring & 1.00 & 0.07 & 0.12 & 15 & 1.00 & 0.07 & 0.13 & 14 & 1.00 & 1.00 & 1.00 & 1 & 0.50 & 0.07 & 0.12 & 14\\
	Open Source management & 0.56 & 0.29 & 0.39 & 78 & 0.51 & 0.32 & 0.39 & 75 & 0.25 & 0.20 & 0.22 & 5 & 0.55 & 0.29 & 0.38 & 75\\
	\rowcolor{lightgray}
	Project management & 0.67 & 0.38 & 0.48 & 77 & 0.60 & 0.40 & 0.48 & 75 & 0.20 & 0.25 & 0.22 & 4& 0.62 & 0.32 & 0.42 & 75\\
	Publishing & 0.64 & 0.46 & 0.54 & 149 & 0.74 & 0.54 & 0.62 & 145 & 0.50 & 0.21 & 0.30 & 19 & 0.67 & 0.54 & 0.60 & 145\\
	\rowcolor{lightgray}
	Reporting & 0.53 & 0.44 & 0.48 & 36 & 0.69 & 0.50 & 0.58 & 36 & 1.00 & 0.50 & 0.67 & 2 & 0.68 & 0.53 & 0.59 & 36\\
	Security & 0.62 & 0.35 & 0.45 & 37 & 0.67 & 0.35 & 0.46 & 34 & 1.00 & 0.33 & 0.50 & 3 & 0.60 & 0.35 & 0.44 & 34\\
	\rowcolor{lightgray}
	Support & 1.00 & 0.08 & 0.14 & 13 & 1.00 & 0.08 & 0.15 & 12 & 1.00 & 1.00 & 1.00 & 1 & 1.00 & 0.08 & 0.15 & 12\\
	Testing & 0.80 & 0.77 & 0.79 & 102 & 0.80 & 0.80 & 0.80 & 100 & 0.50 & 0.20 & 0.29 & 5 & 0.89 & 0.80 & 0.84 & 100\\
	\rowcolor{lightgray}
	Time tracking & 1.00 & 0.50 & 0.67 & 2 & 1.00 & 0.50 & 0.67 & 2 & 1.00 & 1.00 & 1.00 & 1 & 1.00 & 0.50 & 0.67 & 2\\
	Utilities & 0.85 & 0.91 & 0.88 & 402 & 0.86 & 0.92 & 0.89 & 383 & 0.38 & 0.48 & 0.43 & 31 & 0.86 & 0.91 & 0.88 & 383\\ \hline
%	\midrule
	micro avg & 0.79 & 0.71 & 0.75 & 2,053 & 0.80 & 0.71 & 0.75 & 1,956 & 0.48 & 0.49 & 0.49 & 161 & 0.80 & 0.71 & 0.75 & 1,956\\
	macro avg & 0.85 & 0.40 & 0.48 & 2,053 & 0.84 & 0.40 & 0.49 & 1,956 & 0.80 & 0.71 & 0.72 & 161 & 0.85 & 0.41 & 0.49 & 1,956\\
	weighted avg & 0.79 & 0.71 & 0.72 & 2,053 & 0.79 & 0.71 & 0.72 & 1,956 & 0.55 & 0.49 & 0.48 & 161 & 0.79 & 0.71 & 0.72 & 1,956\\
	avg & 0.72 & 0.62 & 0.64 & 2,053 & 0.72 & 0.62 & 0.65 & 1,956 & 0.39 & 0.33 & 0.34 & 161 & 0.73 & 0.63 & 0.65 & 1,956\\ \hline
\end{tabular}
\vspace{-.2cm}
\end{table*}

This section reports the results by answering the %study 
research questions.
%and analyzes the results obtained through the SLR and experiments. We also discuss %about 
%the applicability of \GV, and %as well as 
%the threats that may affect the findings.

\subsection{\rqfirst}
\label{sec:rq1}

We conducted a lightweight SLR~\cite{KITCHENHAM20097}  
inspired by the well-established \emph{``four W-question strategy''} \cite{10.1016/j.infsof.2010.12.010} to select studies 
that deal with the categorization of \GH artifacts, explained as follows.

\smallskip
\noindent
$\triangleright$  
\emph{Which?} We ran a comprehensive search, combining both automated and manual methods to retrieve relevant work. 
The search string used to execute on the title and abstract of the selected papers is as follows: \api{(workflow* OR action* OR CI/CD OR continuous integration OR DevOps OR CICD OR YAML) AND (automat* OR classification OR recommend* OR complet* OR categorization OR tag*) AND (GitHub)}.

\smallskip
\noindent
$\triangleright$
\emph{Where?} Our %literature 
analysis focused on prominent SE %software engineering 
venues, encompassing 9 conferences: ASE, ESEC/FSE, ESEM, ICSE, ICSME, ICST, ISSTA, MSR, SANER, as well as 5 journals: EMSE, IST, JSS, TOSEM, and TSE~\cite{GavelReplicationPackage}. %\footnote{Due to space limits, the acronyms and their full names are listed in the appendix.} 
We accessed the \textit{Scopus} database\footnote{\url{https://scopus.com}} and employed the advanced search and export functions to retrieve all papers published in the aforementioned venues within the given temporal range. In addition, we use a tailored Python script to filter out papers irrelevant to our study~\cite{GavelReplicationPackage}.%\footnote{The script, together with its retrieved results, is available in our replication package.} 

\smallskip
\noindent
$\triangleright$ 
\emph{What?} For each article, we extracted information from the title and abstract by applying predefined keywords to ensure relevance to our research focus.

\smallskip
\noindent
$\triangleright$ 
\emph{When?} %Given that 
As the introduction of GitHub workflows is very recent, 
our search was limited to the most recent five years, \ie 
from 2017 to 2023. This 
allows us to capture the latest developments. 
The query was executed in May 2024, hence 2024 was not considered.

%
%  \begin{lstlisting}[label=lst:searchString,caption=The search string.,style=searchstringstyle,captionpos=t]
	%(workflow* OR action* OR CI/CD OR continuous integration OR 
	%DevOps OR CICD OR YAML) AND (automat* OR classification OR
	%recommend* OR complet* OR categorization OR tag*) AND (GitHub)
	%  \end{lstlisting}

Running the query, we obtained 48 
papers that matched the 
keywords. Afterward, we carefully read the title and abstract, and identified seven studies relevant to the topic under consideration. We filtered out two papers since they analyze generic CI/CD tools, ending up with five papers, 
summarized as follows. 

Valenzuela-Toledo \etal \cite{valenzuela-toledo_evolution_2022} performed a manual categorization of \GA using the card sorting methodology~\cite{10.1007/978-3-031-35132-7_5}. To this end, a dataset of 222 commits and 10 open-source projects has been collected to investigate file modifications. The authors eventually report 11 types of modification in workflows, including primary and secondary groups. The findings 
demonstrate that developers need more support in editing operations of existing workflow.

Calefato \etal \cite{calefato_preliminary_2022} investigated the usage of MLOps operations in %\GH 
workflows. A dataset of 337 valid workflows was extracted from	 existing work. %In particular, 
The authors selected workflows using at least an ML model in the pipeline. After the data collection phase, the dataset is used to search for the most frequent MLOps operations employed. %in the examined workflows. 
%Notably, 
The study reveals that MLOps practices are not different from traditional software systems, apart from specific goals or use cases.

Golzadeh \etal \cite{golzadeh_rise_2022} carried out a large-scale empirical study on 8 popular CI services, including \GA. By using an open-source npm API,  91,810 different \GH repositories were analyzed in terms of CI evolution, combination, and popularity. The results show that usage of \GH actions in the CI/CD pipeline is increasing since they provide an easy-to-use integration mechanism. The same authors ran a qualitative study \cite{rostami_mazrae_usage_2023} by interviewing 22 experienced software engineers in CI development. The study paves the way for reusable workflows using GHA. Similarly, Kinsman \etal \cite{kinsman_how_2021} investigated how developers use GHA to automate 
various tasks. 
The study identified five main categories, \ie Continuous Integration, Utilities, Deployment, Publishing, and Code Quality.

%======================
%Overall, we see that the automatic categorization of \GH 
%actions has not been covered yet by the existing literature. % in Software Engineering. 
%This motivates us to conceive an automated technique to assist developers in generating proper categories for an action, given that they already composed a comprehensive \RM file.
%======================

%selecting the proper actions given their application domain.

%\input{src/Phuong}
%\input{src/Claudio}
%\input{src/Juri}

%\abox{\textbf{Answer to RQ$_1$:}  The lightweight SLR reveals that the usage of \GH actions is still in its infancy, 
%	and no automated technique to classify actions into categories has been proposed yet, triggering the need for a suitable tool.}

\vspace{.2cm}
\noindent\fbox{\begin{minipage}{0.98\columnwidth}
		\paragraph{\textbf{Answer to RQ$_1$:}}  The lightweight SLR reveals that the usage of \GH actions is still in its infancy, 
		and no automated technique to classify actions into categories has been proposed yet, triggering the need for a suitable tool.
\end{minipage}}

\begin{figure*}[t!]
	\centering
	\begin{tabular}{c c c}	
		\subfigure[Precision]{\label{fig:Precision}
			\includegraphics[width=0.26\linewidth]{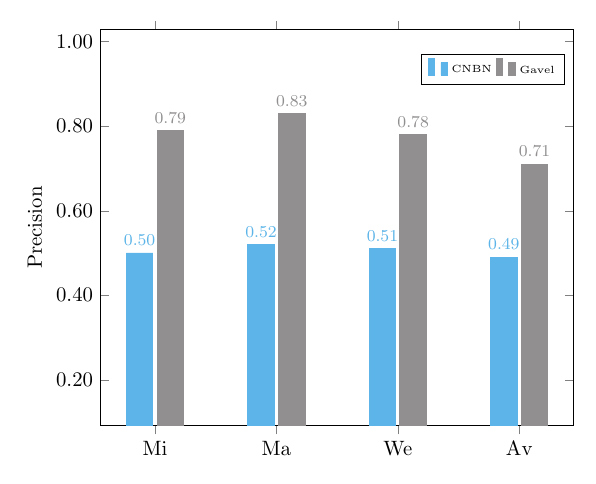}}	&
		\subfigure[Recall]{\label{fig:Recall}
			\includegraphics[width=0.26\linewidth]{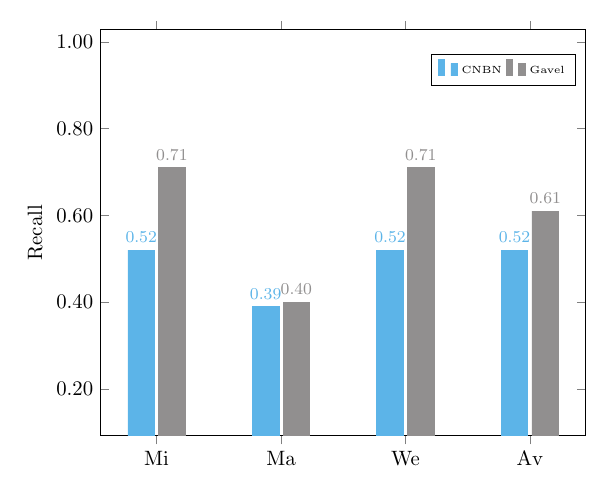}}    &
		\subfigure[F$_1$ score]{\label{fig:F1}
			\includegraphics[width=0.26\linewidth]{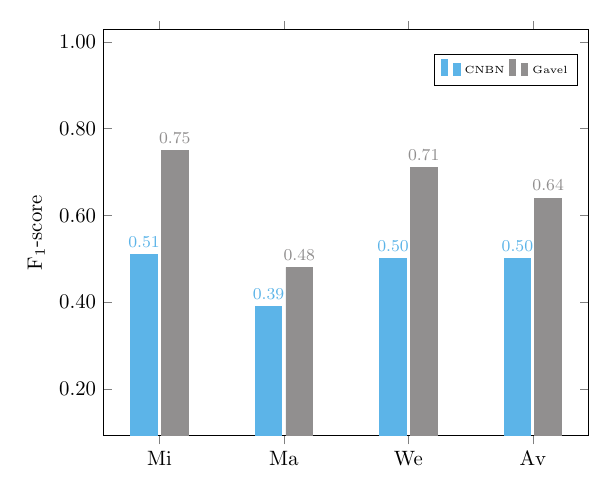}}
	\end{tabular} 
	\vspace{-.2cm}
	\caption{Comparison of accuracy between \CNB and \GV.} 	% different levels of details
	\label{fig:Comparison_CNBN_Gavel}
	\vspace{-.4cm}
\end{figure*}

\subsection{\rqsecond}

We experimented \GV with different configurations shown in Table~\ref{tab:Configurations}. 
Running the evaluation following the settings presented in Section~\ref{sec:Evaluation}, we obtained a series of experimental results. By carefully investigating them, we noticed that by C$_5$, \GV gets a comparable performance compared to that when using C$_1$, C$_2$, and C$_4$, thus we do not depict the results for C$_5$ in the table due to space limits, but report them later in the comparison with \CNB in RQ$_3$.

Table~\ref{table:classification_all} depicts the 
results for the four configurations. 
By each category, we show Precision, Recall, F$_1$ score together with the number of samples 
(\#). It is evident that the categories are considerably imbalanced, \ie while some of them have a large number of samples, most of the categories are sparse.
Investigating the table, we see that \GV obtains perfect precision by various categories. However, 
the recall scores are low in some of the categories.  
While high precision means that the recommendations are mostly relevant and accurate, low recalls indicate that the system fails to recommend many of the relevant items.

With C$_1$, C$_2$, and C$_4$, the 
performance of \GV is good by categories in which there is a large number of samples. 
For example, by \texttt{Utilities} where there are more than 400 samples (C$_1$), the precision, recall, and F1 scores are 0.85, 0.91, and 0.88, respectively. Similarly, with \texttt{Continuous integration}, the 
scores are always greater than 0.85, implying a satisfying accuracy. 
With C$_3$, \GV performs worse 
by various categories, \eg F1 is 0.45 by \texttt{Deployment}. However,  also with C$_3$, 
\GV gets a good prediction by sparse categories. For instance, we count 16 among 30 categories by which there is just one testing sample, \eg 
\texttt{Desktop tools}, \texttt{GitHub Sponsors}, \texttt{IDEs}, but \GV can properly classify them to the right class. Nevertheless, when summing up with the micro, macro, and weighted average values, we see that the accuracy obtained with  C$_3$ is very low, \ie only macro average is greater than 0.70, and the remaining scores are lower than 0.50. %In fact, this is the configuration with the smallest number of samples (see Table~\ref{tab:Configurations}), and such a lack of data impacts on the prediction. 

%\abox{\textbf{Answer to RQ$_2$:} When only code comments 
%	are considered as input, \GV obtains the worst performance. Meanwhile, its prediction 
%	is comparable by the remaining configurations.}

\vspace{.2cm}
\noindent\fbox{\begin{minipage}{0.98\columnwidth}
		\paragraph{\textbf{Answer to RQ$_2$:}} When only code comments 
		are considered as input, \GV obtains the worst performance. Meanwhile, its prediction 
		is comparable by the remaining configurations.
\end{minipage}}

\subsection{\rqthird}

%As we found out through 
RQ$_1$ shows that   
there has been no approach conceived to the categorization of \GH actions. Therefore, we have to find a comparable approach for benchmarking \GV. By investigating the related work in Section~\ref{sec:RelatedWork}, we noticed that \CNB~\cite{di_rocco_hybridrec_2022} is a state-of-the-art tool that tackles a similar issue, \ie 
supporting multi-label categorization of generic \GH projects using \RM files. %
%In addition, the selected model is tailored to handling unbalanced data \cite{10.5555/3041838.3041916}.  
%Thus, the tool has been chosen for the comparison with \GV. 
To avoid any bias, 
we select the best configuration of \GV as well as of \CNB, \ie %. In particular, 
with \GV we used the results obtained for 
C$_5$, corresponding to a slightly better accuracy compared with the remaining configurations.
%Moreover, 
We set the \CNB cutoff value to 2, since it represents the median value of the considered GHA dataset. 

Figure \ref{fig:Comparison_CNBN_Gavel} depicts the comparison between \GV and \CNB in terms of 
Precision, Recall, and F1 score. For each metric, we report its micro average (Mi), macro average (Ma), weighted average (We), and Average (Av) scores.
Notably, \GV outperforms \CNB in all the 
metrics, \eg %. 
%For instance, 
concerning Precision, the corresponding scores for \CNB are always lower than 0.53, while \GV obtains much higher values, greater than 0.70. The same can also be seen with the other metrics. %, \ie Recall and F1-score. 
Altogether, it is evident that our proposed approach yields a substantially superior accuracy compared to that 
by \CNB.

Besides accuracy,
we also measured the time for the training and testing phases of the two 
models. 
Due to space limits, we cannot report the detailed measurement here, but give a brief summary as follows. By examining the results, we can see that compared to \GV, \CNB is much faster both in training and testing. On average, the \CNB model requires around one second for the training phase and less than one second for each fold to perform the testing phase. \GV requires more time for both phases, \ie several hundreds seconds. In this respect, \CNB can be seen as a valuable alternative when there is no hardware resource 
to fine-tune \GV.

\vspace{.1cm}
%<<<<<<< HEAD
%\abox{\textbf{Answer to RQ$_2$:} \GV substantially outperforms \CNB with respect to precision, recall, and F1-score.}
%=======
%\abox{\textbf{Answer to RQ$_3$:} \GV substantially outperforms \CNB in terms of precision, recall, and F1-score. However, \CNB is more timing efficient compared to \GV, 
%	complement to \GV when no powerful platform is available for training.}
	
\vspace{.2cm}
\noindent\fbox{\begin{minipage}{0.98\columnwidth}
		\paragraph{\textbf{Answer to RQ$_3$:}} \GV substantially outperforms \CNB in terms of precision, recall, and F1-score. However, \CNB is more timing efficient compared to \GV, 
		complement to \GV when no powerful platform is available for training.
\end{minipage}}

%>>>>>>> 73cb88e7ebaa16f76e98d560aa0837b7c2903f33

%\begin{figure}[h!]
%	\centering
%	\vspace{-.25cm}
%	\includegraphics[width=0.40\textwidth]{figs/Comparison_Gavel_CNBN_Precision.pdf}
%	\caption{Precision.}
%	\vspace{-.2cm} 
%	\label{fig:Precision}
%	\vspace{-.2cm} 
%\end{figure}
%
%
%\begin{figure}[h!]
%	\centering
%	\vspace{-.25cm}
%	\includegraphics[width=0.40\textwidth]{figs/Comparison_Gavel_CNBN_Recall.pdf}
%	\caption{Recall.}
%	\vspace{-.2cm} 
%	\label{fig:Recall}
%	\vspace{-.2cm} 
%\end{figure}

\subsection{Limitations} 
%\MAX{the part about tokenizer requires clarification, i.e., explain that this is a tokenizer applied before the model's tokenizer} \PN{No, actually we use the tokenizer provided by SetFit, we just explain how it works in Section 3}.

Though \GV was conceived to work with different tokenizers, in the scope of this paper,  
%as a proof of concept, 
we managed to apply a common tokenizer to encode all types of data %\ie 
(code, comments, text). In fact, using a sole tokenizer cannot capture all the intrinsic features in each data type. This can be considered as a limitation of our work, leaving room for improvement. 
%Though \GV is effective, %gains encouraging performance, 
We conjecture that tokenizers tailored to each type of input data will help the tool capture the intrinsic features, thus further improving its ability to provide suitable categories.

%As shown in RQ$_3$, 
\GV obtains a better prediction accuracy compared to that of \CNB. 
However, it suffers from prolonged fine-tuning, though being trained with few-shot samples. 
We see that while pre-trained models are effective, they come at a price in terms of 
training resources. 
%In this respect, 
Thus, conventional models like \CNB can be seen as a valuable alternative when no powerful platforms 
are available for fine-tuning the \GV model.  
In this respect, %Under the circumstances, 
it is necessary to look for more lightweight techniques---compared to transformer---that can maintain a trade-off between accuracy and effectiveness.

\subsection{Threats to Validity}

\smallskip
\noindent
$\triangleright$ \emph{Internal validity}. This is related to the fact that our evaluation might not resemble real-world scenarios, \ie some categories are supported by few samples. To mitigate this, we curated the data collection by retrieving 
information from different sources. The 
SLR may not cover all relevant work, to reduce the threat,
we considered prominent SE venues and carefully checked the titles and abstracts. 
The comparison with \CNB was done using its original implementation.

\smallskip
\noindent
$\triangleright$ \emph{External validity}. This concerns the generalizability of the findings outside this study, \ie \GV may not perform similarly if applied on different datasets. To mitigate this, we experimented with five different configurations considering various kinds of data, \ie plain text, code, or mixed content.

\section{Conclusion and Future work}
\label{sec:Conclusion}
%\section{Conclusion and Future work}

%This paper 
We presented \GV as a practical approach to the categorization of \GH actions. Through a literature review conducted on premier SE venues, we realized that this is the first tool dealing with this issue. An empirical evaluation on a dataset collected from \GH Marketplace demonstrated that our proposed approach can provide suitable categories for actions, and it gains a better performance compared to a state-of-the-art benchmark.
For future work, we plan to equip \GV with lightweight classification models, aiming to maintain a trade-off between effectiveness and efficiency. % %the ability to perform also summarization tasks, in which it produces a short description for a long \RM file.

\begin{acks} 
	
	%All the numerical simulations/evaluations have been realized on the Linux HPC cluster Caliban of the High-Performance Computing Laboratory of the Department of Information Engineering, Computer Science and Mathematics (DISIM) at the University of L'Aquila. 
	This work has been partially supported by the EMELIOT %national 
	research project, which has been funded by the MUR under the PRIN 2020 program grant n. 2020W3A5FY. 
	The work has been also partially supported by the European Union--NextGenerationEU through the Italian Ministry of University and Research, Projects PRIN 2022 PNRR \emph{``FRINGE: context-aware FaiRness engineerING in complex software systEms''} grant n. P2022553SL. We acknowledge the Italian ``PRIN 2022'' project TRex-SE: \emph{``Trustworthy Recommenders for Software Engineers,''} grant n. 2022LKJWHC. %Our research is also funded by Hanoi University of Science and Technology (HUST), Vietnam under project number T2023-PC-002.
\end{acks}

%\section*{Acknowledgement}
%This work has been partially supported by ARS01\_00540 - RASTA  project, funded by the Italian Ministry of Research PNR 2015-2020. The authors also acknowledge the support of the MUR (Italy) Department of Excellence 2023 - 2027 for GSSI. 

\balance
\bibliographystyle{ACM-Reference-Format}
\bibliography{main}

%%
%% If your work has an appendix, this is the place to put it.

\end{document}